\documentclass[aip,jcp,reprint]{revtex4-1}

\usepackage{graphicx}
\usepackage{dcolumn}
\usepackage{bm}
\usepackage[utf8]{inputenc}
\usepackage[T1]{fontenc}
\usepackage{mathptmx}
\usepackage{etoolbox}
\usepackage{balance}
\usepackage{soul}
\usepackage{graphicx}
\usepackage{mathrsfs}
\usepackage{physics2}
\usepackage{xcolor, soul}
\usepackage{multirow, tabularx, booktabs}
\usepackage{braket}
\usepackage{adjustbox}
\usepackage{tablefootnote}
\usepackage{subcaption}
\usepackage{xr} 
\usepackage{threeparttable}
\usephysicsmodule{braket}
\usepackage{placeins}

\usepackage[version=4]{mhchem}
\usepackage{physics}
\usepackage{amsmath}

\newcommand{\Lower}[1]{\smash{\lower 1.5ex \hbox{#1}}}
\makeatletter
\def\@email#1#2{%
 \endgroup
 \patchcmd{\titleblock@produce}
  {\frontmatter@RRAPformat}
  {\frontmatter@RRAPformat{\produce@RRAP{*#1\href{mailto:#2}{#2}}}\frontmatter@RRAPformat}
  {}{}
}
\makeatother

\begin{document}

\preprint{AIP/123-QED}
\title[]
{Frozen density embedding with pCCD electron densities}
\author{Rahul Chakraborty}
\affiliation{Institute of Physics, Faculty of Physics, Astronomy, and Informatics, Nicolaus Copernicus University in Toruń, Grudziądzka 5, 87-100 Toru\'{n}, Poland.}
\author{Pawe\l{} Tecmer}
\affiliation{Institute of Physics, Faculty of Physics, Astronomy, and Informatics, Nicolaus Copernicus University in Toruń, Grudziądzka 5, 87-100 Toru\'{n}, Poland.}
 \email{rahuldev.chakraborty91@gmail.com}

\date{\today}
\begin{abstract}
The pair-coupled-cluster doubles (pCCD) method has emerged as a viable approach for quantum-chemical studies of strongly correlated systems. 
Despite its lower formal scaling ($\mathcal{O}(N^4)$) compared to other versions of coupled cluster (CC) theory, applications to large chemical structures are still expensive.
Fragmentation and embedding strategies offer a viable approach in such cases.
In this work, we present a simple and efficient density-embedding scheme based on pCCD electron densities.
The main computational benefit arises from the fact that pCCD response $\Lambda$-equations are much cheaper to compute than those of standard CC methods, providing easy access to one-electron properties. 
The pCCD densities of the individual subsystems are used to generate static embedding potentials that capture the environment's effect on the embedded system. 
The individual fragment energies are then iteratively converged in a self-consistent fashion.
We demonstrate the reliable performance of this scheme with the estimation of dipole moments of the weakly bound \ce{CO2\cdots Rg} [Rg = He, Ne, Ar, and Kr] complexes and with the modeling vertical excitations of some microsolvated molecules.

\end{abstract}

\maketitle
\begin{quotation}
\end{quotation}
\section{\label{sec:introduction}Introduction}
Quantum chemical studies of molecules and complexes mainly rely on an accurate description of the interplay of interactions between atoms or molecular fragments. 
However, even the most efficient electronic structure algorithms available today cannot handle large structures due to computational demands. 
In the condensed phase, much of the chemistry occurs in a small part of the structure (the so-called active region or active subsystem), and the environment plays a key role in the energetics through its interactions with the active part. 
Keeping this in mind, the problem's complexity can be reduced by using a fragmentation approach.~\cite{gordon2012fragmentation, gomes-emb-review-apsc-2012, jones2020embedding, fragmentation-review-wires-2023}
Quantum embedding theories have made significant strides in this direction.
One class of such embedding theories is where both active and environment fragments are treated with the same or different quantum-mechanical (QM) theories, i.e., QM/QM embedding.~\cite{sun2016embedding}
In this aspect, Cortona introduced the frozen density embedding (FDE) scheme for solid states,~\cite{cortona1992direct} and Wesołowski and Warshel for solvated molecules.~\cite{wesolowski1993fde}
In its original formulation, FDE is a Density Functional Theory (DFT)-based approach~\cite{wesolowski2015fde} where the total density of the system is partitioned into smaller, coupled subsystems, 

\begin{equation}
    \rho_{\textrm{tot}}=\rho_{I} + \rho_{II}.
\end{equation}

Here $\rho_{\textrm{tot}}$ is the supramolecular density; $\rho_{I}$ is the density of the active fragment/subsystem of our interest where the chemistry is happening, and  $\rho_{{II}}$ represents the density of the `frozen' environment.
The supramolecular energy can then be expressed as a sum of fragment energies plus an additional coupling term---$E_{\textrm{int}}[\rho_{I}, \rho_{II}]$,

\begin{align}
\label{eq:E_embed}
    E_{\textrm{tot}}[\rho_{I} + \rho_{II}] = E_{I}[\rho_{I}] + E_{II}[\rho_{II}] + E_{\textrm{int}}[\rho_{I}, \rho_{II}].
\end{align}

The individual subsystem energy can be expressed as

\begin{align}
\label{eq:E_i}
E_{i}[\rho_{i}] 
&= \int \rho_{i}(\mathbf{r}) \, v_{\textrm{nuc}}^{i}(\mathbf{r}) \, d\mathbf{r}
+ \frac{1}{2} \iint \frac{\rho_{i}(\mathbf{r}) \, \rho_{i}(\mathbf{r}')}{|\mathbf{r} - \mathbf{r}'|} \, d\mathbf{r} \, d\mathbf{r}' \nonumber \\
&\quad + V_{\textrm{nuc,nuc}}^{i}
+ E_{\textrm{xc}}[\rho_{i}]
+ T_{\textrm{s}}[\rho_{i}]
\end{align}

where $\nu^{i}_{\textrm{nuc}}$ in the first term is the nuclear charge, the second term denotes the Coulomb potential, and V$^{i}_{\textrm{nuc, nuc}}$ is the nuclear repulsion energy of subsystem $i$ ($i=I, II$).
In Eq.~\eqref{eq:E_i}, $E_{\textrm {xc}}$ and $T_{\textrm{s}}$ denote the exchange--correlation (xc) energy and kinetic energy of the individual subsystem(s), respectively.
The $E_{\textrm{int}}[\rho_{I}, \rho_{II}]$ coupling term in Eq.~\eqref{eq:E_embed} represents the interaction between the subsystems I and II, and depends on their Kohn--Sham electron densities and nuclear charges,

\begin{align}
\label{eq:E_int}
    E_{\textrm{int}}[\rho_{I}, \rho_{II}] &= \int \rho_{I}(\mathbf{r}) v_{\textrm{nuc}}^{II}(\mathbf{r}) d\mathbf{r} + \int \rho_{II}(\mathbf{r}) v_{\textrm{nuc}}^{I}(\mathbf{r}) d\mathbf{r} \nonumber\\ 
    &+ \iint \frac{\rho_{I}(\mathbf{r}_1) \rho_{II}(\mathbf{r}_2)}{|\mathbf{r}_1 - \mathbf{r}_2|} d\mathbf{r}_1 d\mathbf{r}_2 + V^{I-II}_{\textrm{nuc, nuc}} \nonumber \\
    &+ E_{\textrm{xc}}^{\textrm{nadd}}[\rho_{I}, \rho_{II}] + T_{\textrm{s}}^{\textrm{nadd}}[\rho_{I}, \rho_{II}].
\end{align}

The non-additive energy contributions coming from the xc and kinetic terms are defined as

\begin{equation}\label{eq:xc_nadd}
    E_{\textrm {xc}}^{\textrm{nadd}}[\rho_{I}, \rho_{II}]=E_{\textrm {xc}}[\rho_{I}+\rho_{II}] - E_{\textrm {xc}}[\rho_{I}] -E_{\textrm {xc}}[\rho_{II}]
\end{equation}
and
\begin{equation}\label{eq:s_nadd}
    T_{s}^{\textrm{nadd}}[\rho_{I}, \rho_{II}]=T_{\textrm {s}}[\rho_{I}+\rho_{II}] - T_{\textrm {s}}[\rho_{I}] -T_{s}[\rho_{II}],
\end{equation}

respectively.
The quantities in Eqs.~\eqref{eq:xc_nadd} and~\eqref{eq:s_nadd} are estimated using approximate density functionals.
The computational advantage of expressing the energy of the whole system according to Eq.~\eqref{eq:E_embed} arises from the fact that only one subsystem is optimized self-consistently at a time, keeping the electron density of the other subsystem `frozen'. 
The derivative of the interaction energy functional that describes the inter-system dependence is the embedding potential, which can be decomposed into Coulomb interactions with the environment (nuclei and frozen electron density) plus derivatives of the non-additive parts of the xc and kinetic energy functionals.
The performance of the FDE approach heavily relies on the non-additive xc and kinetic terms in the employed density functionals.~\cite{jacob2007exact, fux2010accurate}
Numerical results show that the latter term is the more dominating one, especially for strongly interacting (e.g., via covalent bonds) subsystems, and acts as a major source of error in FDE calculations.~\cite{gotz2009performance, goodpaster2010exact, fscc-cuo-noble-gases-emb-jcp-2012}  
Substantial work on excitation energy and response property calculations using time-dependent density functional theorem (TD-DFT) within the FDE framework has been done by different groups, which have expanded the applicability of FDE.~\cite{casida2004generalization, neugebauer2007couplings, neugebauer2009response}
Projection-based embedding techniques~\cite{manby2012simple, lee2019projection, beran2023projection, monino2026projection} and their Huzinaga equation-based versions~\cite{hegely2016exact, graham2020robust, csoka2024development} constitute a path of development parallel to density embedding methods.

Due to the flexibility in the nature of the density used in FDE, wave-function theory (WFT)-based approaches can be easily integrated into this framework. 
The WFT-in-DFT version of the embedding approach employs various WFT methods for the `active' subsystem while the environment is treated at the DFT level. 
The formal correctness of such an approach, for exact density functional along with full configuration interaction (FCI), was shown by Wesołowski.~\cite{wesolowski2008embedding}
Carter and coworkers have done extensive work using variational WFT methods, such as configuration interaction,~\cite{govind1998accurate} complete active space self-consistent field (CASSCF),~\cite{govind1999electronic} to study localized excitations in solids or surfaces.
Coupled cluster (CC) methods have also been used in this context.
H{\"o}fener et al. implemented a CC-in-DFT method based on the response formalism.~\cite{hofener2012molecular, hofener2013solvatochromic}
A simplified approach was explored by Gomes and coworkers, in which a DFT-in-DFT FDE calculation is performed to obtain converged subsystem densities and the resulting embedding potential, which is then used to represent the environment in the WFT-in-DFT calculation.~\cite{gomes2008calculation, fscc-cuo-noble-gases-emb-jcp-2012, gomes2013towards}
Recent works have also explored WFT-in-WFT frozen density embedding, thereby alleviating dependence on DFT and its associated problems.~\cite{hofener2012_cc-in-cc, hofener2016wave}

pCCD-based methods have attracted significant attention as efficient and viable WFT computational models in recent years.
Though tracing its origin to geminal theories,~\cite{limacher-ap1rog-jctc-2013} pCCD can also be considered as a simplification of the CC ansatz,~\cite{tamar-pccd-jcp-2014} which includes only paired excitations from the same spatial orbitals.
pCCD has demonstrated success in treating complex electronic structures at a mean-field computational cost.~\cite{oo-ap1rog-prb-2014, ps2-ap1rog-jcp-2014, henderson2014seniority, ap1rog-non-variational-orbital-optimizarion-jctc-2014,  pccd-correlation-analysis-prb-2016, pccp-geminal-review-pccp-2022}
The method has been utilized to model electronic structures of a range of chemical species, such as actinides,~\cite{ap1rog-singlet-gs-actinides-pccp-2015} organic molecules,~\cite{pccd-perspective-jpcl-2023, eom-pccd+s-ct-analysis-carbazole-ram-jpca-2025} studying catalytic action of molybdenum cofactor,~\cite{pccd-mocco-galynska-pccp-2024} etc.
A plethora of recent works have used pCCD to model excited states.~\cite{pccd-ee-f0-actinides-pccp-2019, lr-pccd-jctc-2024, benchmark_ip_pccd, dip-fpccsd-jctc-2025, eom-pccd+s-ct-analysis-carbazole-ram-jpca-2025}  
However, simulating larger structures, especially in condensed phases, remains a challenge.
This is further corroborated by the need for post-pCCD dynamic electron-correlation corrections to achieve chemical accuracy, which steeply increases the computational demand.
This motivates the development of pCCD-based embedding techniques.

In our previous work, we implemented a pCCD-in-DFT approach,~\cite{pccd-static-embedding} where the static embedding potential was generated from DFT-in-DFT calculations in external software.
In this work, we focus on the implementation and analysis of the WFT-in-WFT computational scheme based solely on electron densities obtained from pCCD. 
The key advantage over conventional CC-in-CC embedding approaches is that the pCCD one-particle density matrix can be obtained straightforwardly by solving the associated $\Lambda$-equations, which scale as $\mathcal{O}(N^4)$.
This results in a simple, fully flexible, and computationally efficient pCCD-based embedding framework that introduces virtually no additional cost beyond the underlying pCCD calculation.
The performance of the method is assessed on two test cases: (i) the dipole moments of weakly bound \ce{CO2}$\cdots$Rg [Rg = He, Ne, Ar, and Kr] complexes and (ii) the vertical excitation energies of selected microsolvated systems.

This work is structured as follows: Section~\ref{sec:theory} offers a concise overview of the theoretical models under investigation, followed by a detailed description of the computational methods in Section~\ref{sec:comput-det}.
Section~\ref{sec:results} summarizes the numerical results.
Concluding remarks are presented in Section~\ref{sec:conclusions}.
\section{Theory}\label{sec:theory}
\subsection{Ground and excited states with pCCD-based methods}
The pCCD ansatz was inspired by the concept of electron-pair wave functions (geminals~\cite{limacher-ap1rog-jctc-2013}) and originally developed as the antisymmetric product of a one-reference geminal model. 
It can be expressed as a simple form of the CC ansatz~\cite{tamar-pccd-jcp-2014} 

\begin{equation}
\label{eqn:equation-pccd}
     \Psi_{\textrm{pCCD}} = e^{\hat{T}_{\textrm{p}}} \ket{\Phi_0},
\end{equation}   

where

\begin{equation}
\label{eqn:equation-t2p}
     \hat{T}_\textrm{{p}} = \sum_{i}^{n_{\textrm {occ}}} \sum_{a}^{n_{\textrm {virt}}} 
    t^{a\bar{a}}_{i\bar{i}} \hat{a}^\dagger_{a} \hat{a}^\dagger_{\bar{a}} \hat{a}_{\bar{i}} \hat{a}_{i}
\end{equation}

is a cluster operator that excites electron pairs from the same spatial orbital.
Here ${ \hat{a}^{\dagger}_{p} ({\hat{a}^{\dagger}_{\bar{p}}})}$ are the electron creation and  ${ \hat{a}_{p} ({\hat{a}_{\bar{p}}})}$ are the electron annihilation operators for the alpha ${( p)}$ and beta ${(\bar{p})}$ electrons, respectively.
{${t^{a\bar{a}}_{i\bar{i}}}$} are the pCCD cluster amplitudes, and $\ket{\Phi_0}$ is the reference wave function. 
When augmented with a variational orbital optimization (oo) protocol, pCCD becomes size-consistent.~\cite{oo-ap1rog-prb-2014, diatomics-oo-ap1rog-jpca_2014} 
Several studies have shown that oo-pCCD provides a reliable description of quasi-degenerate states and allows for the modeling of potential energy surfaces.~\cite{oo-ap1rog-prb-2014,  limacher2014influence, ap1rog-non-variational-orbital-optimizarion-jctc-2014,pccd-yb2-ijqc-2019, pccd-non-covalent-intwractions-jctc-2019, pawel-geminal-review-pccp-2022}

However, a major drawback of this model is the lack of description of dynamic electron correlation, as pCCD specifically focuses on the seniority-zero part of the full electronic Hamiltonian.
A remedy for this is to use the tailored CC  (tCC).~\cite{kinoshita2005coupled, tailored-cc-markus-jcp-2020}
In tCC, the excitation space is partitioned~\cite{tailored-cc-implemntation-oliphant-jcp-1992, tailored-cc-jcp-1993, tailored-cc-jcp-1994} using the cluster operators $\hat{T}_{\textrm {int}}$ and $\hat{T}_{\textrm {ext}}$. 

\begin{equation}
     \label{eq:tcc}
     \ket{\Psi_{\textrm{tCC}}} 
        =e^{\hat{T}_{\textrm {ext}}}e^{\hat{T}_{\textrm {int}}}\ket {\Phi_{0}}.
\end{equation}

First, $\hat{T}_{\textrm {int}}$, describing the strong correlation effect, is applied to the
reference wave function $\ket{\Phi_{0}}$. Subsequently, $\hat{T}_{\textrm {ext}}$ is applied to
this intermediate wave function, with the  ${t}_{\textrm {int}}$ amplitudes
kept constant to introduce additional excitations and
incorporate dynamic correlation in the final wave
function. 
In a similar fashion, the dynamic correlation excluded in the
pCCD/oo-pCCD wave function can be accounted for by \textit{a posteriori}
corrections, where single and unpaired double excitations are
added to the wave function. 
Such an approach is termed as fpCC,~\cite{ola-tcc} which
expands the pCCD reference wave function by adding unpaired doubles, along with single
excitations (fpCCSD) to $\ket{\Psi_{\textrm {pCCD}}}$,~\cite{frozen-pccd-jcp-2014}
\begin{equation}
     \label{eq:fpcc}
     \ket{\Psi_{\textrm{fpCCSD}}} 
        =  e^{\hat{T}_{\textrm {ext}}}e^{\hat{T}_{\textrm {p}}}\ket {\Phi_{0}}.
\end{equation}

In Eq.~\eqref{eq:fpcc}, $\hat{T}_{\textrm {ext}}$ is the cluster operator
which acts upon the pCCD wave function and generates non-pair doubles (also known as broken pairs) and singles excitation.
The amplitudes of $\hat{T}_{\textrm {p}}$, which acts as $\hat{T}_{\textrm {int}}$ in this case, are kept constant (and hence the name `frozen pair').
Specifically, in the spin-free CC formulation, where all excitation operators are written in terms of singlet excitation operators $\hat{E}_{{a}i}$, $\hat{T}_{\textrm {ext}}$ reads as

\begin{align}
\label{eqn:T_ext}
{\hat{T}_{\textrm {ext}}= \hat {T}_1 + (\hat{T}_2 -\hat{T}_{\textrm {p}}) = \sum_{i}^{n_{\textrm {occ}}} \sum_{a}^{n_{\textrm {virt}}} t_{i}^{{a}} \hat{E}_{{a} i} + \frac{1}{2}\sum_{i, j}^{n_{\textrm {occ}}}~ \sum_{{a}, b}^{n_{\textrm {virt}}}~t_{ij}^{{a}b}~\hat{E}_{{a}i}\hat{E}_{bj} } \nonumber \\
- \frac{1}{2}\sum_{i}^{n_{\textrm {occ}}}~ \sum_{{a}}^{n_{\textrm {virt}}}~t_{ii}^{{a}a}~\hat{E}_{{a}i}\hat{E}_{ai}, 
\end{align}

where

\begin{equation}
\label{eqn:q_ai}
{\hat{E}_{{a}i} = \hat{a}_{{a}}^{\dagger}\hat{a}_{i} + \hat{a}_{\bar{ a}}^{\dagger}\hat{a}_{\bar{i}}}
\end{equation}   

and $\hat{E}_{{b}j}$ is defined analogously. 
One  simplified form of fpCCSD is the frozen pair linearized coupled cluster singles and doubles (fpLCCSD) ansatz,~\cite{frozen-pccd-jcp-2014, ap1rog-lcc-jctc-2015} 

\begin{equation}
     \label{eq:fplcc}
     \ket{\Psi_{\textrm{fpLCCSD}}} 
        \approx (1+ \hat{T}_{\textrm {ext}})\ket {\Psi_{\textrm{pCCD}}}.
\end{equation}

fpLCC can be understood as a linearized pCCD-tCC approach, where only the linear terms involving $\hat{T}_{\textrm {ext}}$ are included in the amplitude equations.
In that case, the energy is calculated by terminating the Baker–Campbell–Hausdorff expansion at the second term in $\hat{T}_{\textrm{ext}}$,

\begin{equation}
    (\hat{H}_{N}+[\hat{H}_{N},\hat{T}_{\textrm {ext}}]) \ket{\Psi_{\textrm {pCCD}}} = E\ket{\Psi_{\textrm {pCCD}}}.
\end{equation}

Here, $\hat{H}_{N}=\hat{H}-\braket{\Phi_{0}|\hat{H}}{\Phi_{0}}$ is the normal ordered form of the electronic Hamiltonian $\hat{H}$.
This ansatz can be understood as a linearized, pCCD-tailored CC approach, in which the full frozen-pair equations (including all nonlinear and disconnected terms involving $\hat{T}_{\textrm{p}}$ vertices) are included in the LCCSD equations.
Here we would like to point out to the readers that fpLCCSD was also denoted previously as pCCD-LCCSD in the literature.~\cite{ap1rog-lcc-jctc-2015}

Excited states can be modeled with pCCD using the equation-of-motion (EOM) formalism.~\cite{eom-cc-rowe-rmp-1968, eom-cc-cpl-1989, eom-ccsd-bartlett-jcp-1993}
The simplest excited state extension for pCCD is the EOM-pCCD model, where the cluster operator $\hat{T}$ and the CI-type operator $\hat{R}$ are restricted to pair-double excitations only, 

\begin{equation}
\label{eq:EOM-R}
\hat{R}_{\textrm{p}}
= c_{0}\,\hat{\tau}_{0}
+ \sum_{i a} c_{ii}^{aa} \, \hat{\tau}_{ii}^{aa}.
\end{equation}

Here, $\hat{\tau}_{ii}^{aa}=\frac{1}{2}\hat{E}_{{a}i} \hat{E}_{{a}i}$ denotes pair excitations, and $\hat{\tau}_0$ is the identity operator which describes the contribution of the ground state in the particular excited state.
The corresponding EOM-pCCD equation for a given excited state $k$ can be expressed as 

\begin{equation}
\label{eq:EOM-pCCD}
     [\mathscr {\hat{H}}^{\textrm {(p)}}_{N}, \hat{R}_{\textrm {p}}] \rvert \Phi_0 \rangle
= \omega_k  \hat{R}_{\textrm {p}} \rvert \Phi_0 \rangle,
\end{equation}

where 

\begin{align}
\label{eq:H_sim_pccd}
\mathscr{\hat{H}}^{\textrm {(p)}}_{N}
&= e^{-\hat{T}_{\textrm{p}}} ~\hat{H}_{N}~ e^{\hat{T}_{\textrm{p}}} \\ \nonumber
&= \hat{H}_{N}
+ [\hat{H}_{N}, \hat{T}_{\textrm {p}}]
+ \frac{1}{2} [[\hat{H}_{N}, \hat{T}_{\textrm {p}}], \hat{T}_{\textrm {p}}]
\end{align}

is the similarity-transformed pCCD Hamiltonian, and $\omega_k$ is the excitation energy for the particular state. 
For accessing general (non-pair) doubly excited and singly excited states, $\hat{ T}_{2}^{\prime}$ and $\hat{T}_{1}$ are included in the cluster operator.~\cite{eom-fpccsd-dalton-transaction-2026}
In that case, the ground state wave function is fpLCCSD, and the approach is called EOM-fpLCCSD (also denoted as EOM-pCCD-LCCSD in previous articles).~\cite{eom-pccd-lccsd-jctc-2019} 
$\hat{R}$, including the singles and doubles contributions, bears the form

\begin{equation}
    \hat{R}_{\textrm{sd}}
= c_{0}\,\hat{\tau}_{0}
+ \sum_{i {a}} c_{i}^{ {a}}\,\hat{\tau}_{i}^{a}
+\frac{1}{2} \sum_{ia} \sum_{jb} c_{i j}^{{a} b}\,\hat{\tau}_{i j}^{{a} b}.
\end{equation}

Here $\hat{\tau}_{i}^{a}=\hat{E}_{{a}i}$ is the (spin-free) single excitation operator and $\hat{\tau}_{ij}^{ab}=\hat{E}_{{a}i} \hat{E}_{{b}j}$ is the (spin-free) doubles excitation operator.
In this case, the EOM-fpLCCSD equation, again for a given $k$-th excited state, becomes

\begin{equation}
[\mathscr{\hat{H}}_{N}^{\textrm (sd)},~ \hat{R}_{\textrm {sd}}]\ket{\Phi_0} = \omega_{k} \hat{ R}_{\textrm {sd}} \ket{\Phi_0}.
\end{equation}

It is important to note here that the similarity transformed Hamiltonian $\mathscr{\hat{H}}_{N}^{\textrm{(sd)}}$ has different expressions for pair and non-pair excitations.
For non-pair excitations, it has the following form 

\begin{align}
\hat{\mathscr{H}}^{(\textrm{np})}_{N}
&= e^{-(\hat{T}_{\textrm{p}} - \hat{T}_{\textrm{np}})} \, \hat{H}_{N} \, e^{(\hat{T}_{\textrm{p}} + \hat{T}_{\textrm{np}})} \nonumber \\
&\approx \hat{H}_{N} 
+ [\hat{H}_{N}, \hat{T}_{\textrm{np}}]
+ [[\hat{H}_{N}, \hat{T}_{\textrm{np}}], \hat{T}_{\textrm{p}}] \nonumber \\
& +[\hat{H}_N, \hat{T}_{\textrm {p}}] +\frac{1}{2}[[\hat{H}_N, \hat{T}_{\textrm{p}}], \hat{T}_{\textrm{p}}],
\end{align}

because of the exclusion of terms non-linear in $\hat{T}_{\textrm{np}}$ in the linearized cluster operator, while the pair excitations are still obtained with Eq.~\eqref{eq:H_sim_pccd}.

\subsection{FDE with pCCD}\label{theory:FDE}
In our pCCD-in-pCCD FDE scheme, we obtain subsystem electron densities from the pCCD response one-particle reduced density matrices (1-RDMs)~\cite{henderson2014seniority, pccd-prb-2016} of each subsystem defined as

\begin{equation}
\label{eq:response-1-RDM}
    ^{\textrm {pCCD}}\gamma_{q}^{p}= \bra {\Psi}(1+\hat{\Lambda}_{\textrm {p}})\textrm{e}^{-\hat{T}_{\textrm{p}}}{{\hat{a}_{ p}^{\dagger}\hat{a}_{q}}}\textrm{e}^{\hat{T}_{\textrm{p}}}\ket {\Psi},
\end{equation}

where $\hat{\Lambda}_{\textrm {p}}=\sum_{i{a}}\lambda_{{a} \bar{{a}}}^{i \bar{i}}\hat{a}_{{i}}^{\dagger}\hat{a}_{\bar{i}}^{\dagger}\hat{a}_{{{ {a}}}}\hat{a}_{{a}}$ is the electron-pair de-excitation operator. 
It is to be noted that pCCD 1-RDM is diagonal and depicts the occupation numbers in the natural orbital basis.
The 1-RDM is already obtained during the pCCD orbital-optimization step and does not involve any additional computation.

The derivative of the coupling energy term defined in Eq.~\eqref{eq:E_int} 
with respect to the subsystem density gives the embedding potential

\begin{equation}\label{eq:emb-pot}
\begin{split}
 v_{\textrm{emb}}^{I}(\mathbf{r}) 
  =\;& v_{\textrm{nuc}}^{II}(\mathbf{r}) 
  + \int \frac{\rho_{II}(\mathbf{r}^{\prime})}
               {|\mathbf{r}-\mathbf{r}^{\prime}|} 
    \, \textrm{d}\mathbf{r}^{\prime}  \\
 &+ \left.
    \frac{\delta E_{\textrm{xc}}^{\textrm{nadd}}[\rho_{I},\rho_{II}]}
         {\delta \rho(\mathbf{r})}
    \right|_{\rho_{I}}
  + \left.
    \frac{\delta  T_{\textrm{s}}^{\textrm{nadd}}[\rho_{I},\rho_{II}]}
         {\delta \rho(\mathbf{r})}
    \right|_{\rho_{I}},
\end{split}
\end{equation}

where $\rho_{{i}}$ terms denote the electron density of the i-th subsystem obtained from the corresponding pCCD calculation.
There is no exact method to describe the non-additive terms other than using approximate functionals.
In this work, we used a simple Thomas-Fermi model,~\cite{thomas-fermi-mpcpc-1927, thomas-fermi-theories-rmp-1981} which corresponds to the local density approximation (LDA) in DFT, to describe the non-additive kinetic potential

\begin{align}
v_{\textrm {kin}}^{\textrm {nadd}}(\mathbf r) 
&= \frac{\delta T_{\textrm {TF}}^{\textrm {nadd}}[\rho_{I},\rho_{II}]}{\delta \rho_{I}(\mathbf r)}
\nonumber \\
&= \frac{1}{2} (3\pi^2)^{2/3} \times \Big[ (\rho_{I} + \rho_{II})^{2/3} - \rho_{I}^{2/3} \Big].
\end{align}

The non-additive xc term in the embedding potential of Eq.~\eqref{eq:emb-pot} is estimated using the Slater X$\alpha$ potential~\cite{slater-x-alpha-pr-1951}

\begin{align}
v_{\mathrm{xc}}^{\mathrm{nadd}}(\mathbf{r})
&= \frac{\delta E_{X\alpha}^{\mathrm{nadd}}[\rho_{I},\rho_{II}]}{\delta \rho_{I}(\mathbf r)} \nonumber \\ 
&= - \alpha \frac{3}{2} \left( \frac{3}{\pi} \right)^{1/3} 
\times \left[
(\rho_{I} + \rho_{II})^{1/3}
- \rho_{I}^{1/3}
\right].
\end{align}

The value of the Slater exchange parameter $\alpha$ is actually atom-dependent and also varies for the property under question.
For example, Zope et al. estimated the optimal $\alpha$ values for atomization energies of many different atoms.~\cite{zope2005slater}  
For simplicity, in this work, we used $\alpha$=0.66 as originally proposed by Kohn and Sham.~\cite{kohn1965self} 
The LDA approximation for the non-additive kinetic potential term has been used for weakly interacting systems and has given reliable results; however, for strongly overlapping subsystems, LDA performs rather poorly.~\cite{wesolowski2003gradient}
The Slater exchange potential, on the other hand, does not show proper long-range Coulomb decay. 
And of course, using only the Slater exchange potential in the non-additive xc term means the correlation part is missing from the embedding potential.
These simple models definitely leave room for improvement in the pCCD-based FDE framework.
\begin{figure}
\centering
    \includegraphics[width=\linewidth]{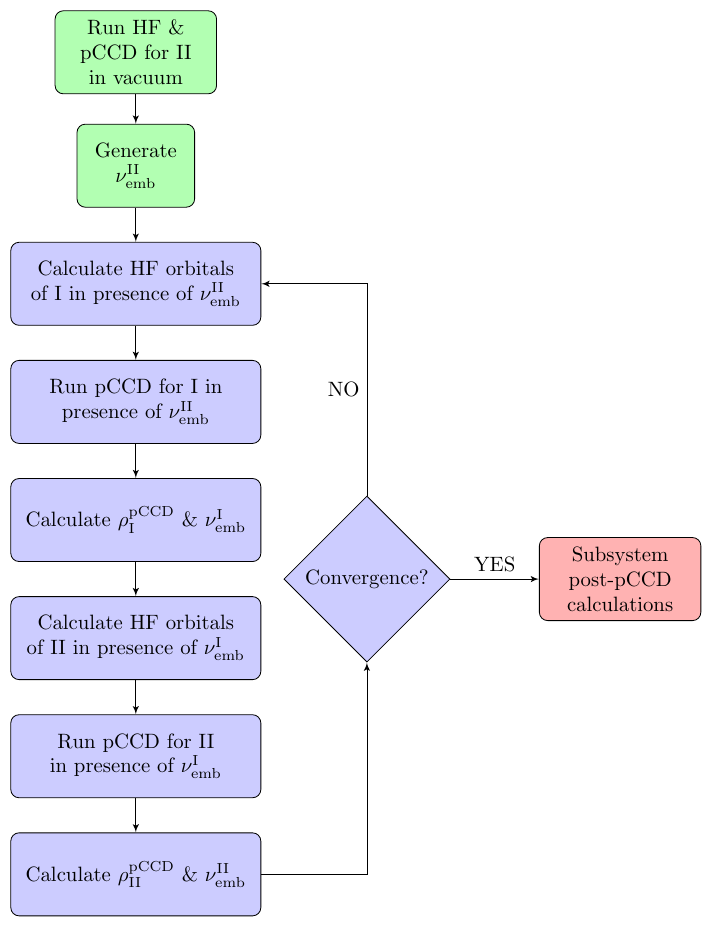}
    \caption{Depiction of FDE freeze-and-thaw cycles with pCCD densities. 
    The green blocks show steps performed with the isolated environment subsystem. The red block shows post-pCCD calculations (fpCCSD, fpLCCSD) with the converged pCCD wave function of each subsystem. 
    EOM-fpLCCSD calculations are also done at this stage after fpLCCSD. }
    \label{fig:fde_flowchart}
\end{figure}

An important aspect to note here is that a completely frozen $\rho_{II}$ can not represent the polarization effect of the environment on the active subsystem and vice versa.
This is solved by employing the so-called `freeze-and-thaw' cycles.~\cite{wesolowski-freeze-and-thaw-cpl-1996}
Here, one subsystem density is calculated at a time, keeping the other subsystem frozen.
Thereafter, the roles of the subsystems are reversed.
The previously frozen subsystem is now thawed, and its density is calculated by freezing the other subsystem's density (obtained in the previous step).
This is done till self-consistency.
Figure~\ref{fig:fde_flowchart} describes the freeze-and-thaw workflow followed in this work. 
At the beginning, the so-called environment subsystem (marked as II) is considered in isolation, i.e., in the absence of the other subsystem (but at the supramolecular geometry). 
HF and subsequent pCCD calculations are run for this fragment.
This works as a guess density to initiate the freeze-and-thaw cycles.
The embedding potential representing the isolated environment subsystem is then evaluated on a numerical grid. 
From the next cycle, the one-electron matrix elements of this potential are calculated as

\begin{equation}
 v_{{ij}}^{{I}}
  = \langle \phi_{{i}} | v_{\textrm{emb}}^{{II}} | \phi_{{j}} \rangle 
 \approx \sum_{{k}} w_{{k}} \, v_{\textrm{emb}}^{{I}}(\boldsymbol{r}_{{k}})\,
  \varphi_{{i}}(\boldsymbol{r}_{{k}})\, \varphi_{j}(\boldsymbol{r}_{k}),
\end{equation}

where $\phi_{i}(r_{k})$ is the numerical value of the orbital $\phi_{i}$ at grid point r$_{k}$  and $w_{{k}}$ is the integration weight of this point.
This term is then plugged into the Fock matrix of the current active subsystem (I) at the HF level, and the SCF iterations are run.
This brings, to a certain extent, the relaxation effect of the orbitals of subsystem I due to the presence of subsystem II.
The resultant perturbed canonical MOs are used for pCCD calculations in the presence of the static embedding potential, which represents the environment/frozen subsystem (II). 
From the updated pCCD electron density of subsystem I obtained at this stage, its embedding potential is generated again.
After that, the roles of the active and environment subsystems are swapped, and the same steps are repeated.
This completes one freeze-and-thaw macroiteration.
These steps are followed until convergence of the pCCD energy of each subsystem.

Post-convergence, fpCCSD/fpLCCSD calculations are run for each fragment using its converged pCCD wave function.
In the case of excitation energy, EOM-fpLCCSD calculations are performed after the fpLCCSD step to get the vertical excitation energies.
A limitation of any static embedding approach, such as ours, is the neglect of the change in electron densities in response to electronic excitations.
Consequently, we always consider a fixed embedding potential based on the ground-state subsystem density.
This can only produce reliable results as long as the non-additive kinetic and xc potentials are not very different between the ground and excited states.
For the weakly bound systems investigated in this work, we assume this to be true.
The effect of electronic excitations in different subsystems can be incorporated following Neugebauer et al.,~\cite{neugebauer2007couplings} which has also been implemented for CC2 by H\"ofener et al.~\cite{hofener2016wave}
\section{Computational details}\label{sec:comput-det}
\subsection{Structures}
Bond parameters of the \ce{CO2 \cdots Rg} complexes were taken from Maroulis et al.~\cite{maroulis2001interaction}
Coordinates of \ce{NH3 \cdots H2O} complex was taken from our previous work.~\cite{pccd-static-embedding}
The solvated uracil structure was taken from Etinski and Marian.~\cite{etinski2010ab}
The xyz structures of all investigated molecules are listed in the SI. 

\subsection{Generation of embedding potential}
The `Grid' package was used and interfaced with PyBEST to generate numerical integration grids for embedding potentials.~\cite{tehrani2024grid}
All embedding calculations involve supramolecular grids.
The grids were constructed using the Becke radial transformation in combination with Lebedev–Laikov angular grids with 131 points for all heavy atoms.~\cite{lebedev1999quadrature}
Becke partitioning weights were used in the numerical integration.

pCCD-based freeze-and-thaw (FT) cycles were run as described in section~\ref{theory:FDE}.
Convergence was accepted when the change in subsystem pCCD energies of both subsystems went below 10$^{-8}$ E$_{\textrm{h}}$ between two consecutive cycles.
For the weakly interacting systems investigated in this work, it takes at most 3--4 FT cycles to fully converge the pCCD energies of each subsystem.
Also, for any subsystem, beyond the step in the FDE workflow (see Figure~\ref{fig:fde_flowchart}) where the effect of the appropriate embedding potential is introduced in the calculation for the first time, the energy change and changes in dipole moment are minimal.
Taking this into account, we also explored another simplified strategy.
The pCCD densities and embedding potentials of the environment subsystem were generated in vacuum; these potentials were then used to embed the active subsystem in a single-step calculation. 
Such an approach is referred to as `non-FT' throughout this work.
\subsection{Coupled cluster calculations in PyBEST}
All pCCD and post-pCCD calculations have been performed in a developer version of the PyBEST v2.1.0dev0 software package.~\cite{pybest-paper-cpc-2021, pybest-paper-update1-cpc-2024, pybest-gpu-jctc-2024}
In this work, we used monomer-centered basis functions, specifically the aug-cc-pVDZ basis set.~\cite{aug-cc-pvxz-jcp-1992} 
Pipek--Mezey~\cite{pipek1989fast} localized orbitals were used in all cases.
However, pCCD orbital optimization was not performed.
The major reason is to avoid computationally expensive orbital rotations during each FT macro-iteration.
As far as the ground state molecular properties are concerned, our previous works have shown that the use of optimized orbitals actually worsens the performance of pCCD-based methods to a certain extent.~\cite{pccd-dipole-moments-jctc-2024, pccd-expectation-value-1dm-jpca-2025}
This is also true for a certain type of excited states.~\cite{lr-pccd-jctc-2024}
In addition, the excitation energies of EOM-fpLCCSD do not differ much between the canonical HF and natural pCCD orbitals. Still, excited-state identification is simpler and easier in the canonical picture.~\cite{eom-pccd-lccsd-jctc-2019}
Furthermore, the behavior of the pCCD orbital optimization during FT cycles needs to be investigated in detail, which is beyond the scope of this work.
Since we are not performing orbital optimization, the reference wave function $\ket{\Phi_0}$ in Eq.~\eqref{eqn:equation-pccd} refers to the HF Slater determinant.
All electronic integrals were generated with the Libint package.~\cite{libint2}
Cholesky decomposition~\cite{cholesky-koch-jcp-2003, cholesky-review-2011} was used with a tight threshold of 10$^{-8}$.
The frozen-core approximation was used in all correlated calculations as implemented in PyBEST.

Expectation-value-based (denoted by XpCCD, XfpCCSD, XfpLCCSD, XCCSD) dipole moments were calculated using the cluster operator amplitudes and $\hat{S}$-operator based approximations,~\cite{korona2011local} as described in our previous publication.~\cite{pccd-expectation-value-1dm-jpca-2025}
The response fpLCCSD dipole moments (marked without X), on the other hand, are calculated by solving the associated $\Lambda$-equations.~\cite{pccd-dipole-moments-jctc-2024} 
All fpCCSD and fpLCCSD calculations are performed after convergence of the freeze-and-thaw cycles in the FT framework.
Though the CCSD calculation is not a post-pCCD step in a theoretical sense, we use the perturbed HF MOs of the subsystems from the final FT iteration, and hence these were also included
as a part of our `post-pCCD' step in Fig.~\ref{fig:fde_flowchart}. 
CCSD(T) response dipole moments were obtained from the Molpro package.~\cite{werner2012molpro, molpro2020_jcp}
\section{Results and discussion}\label{sec:results}
\subsection{Dipole moments of \ce{CO2\cdots Rg} complexes}

\begin{figure}[h!]
    \centering
    \includegraphics[width=\linewidth]{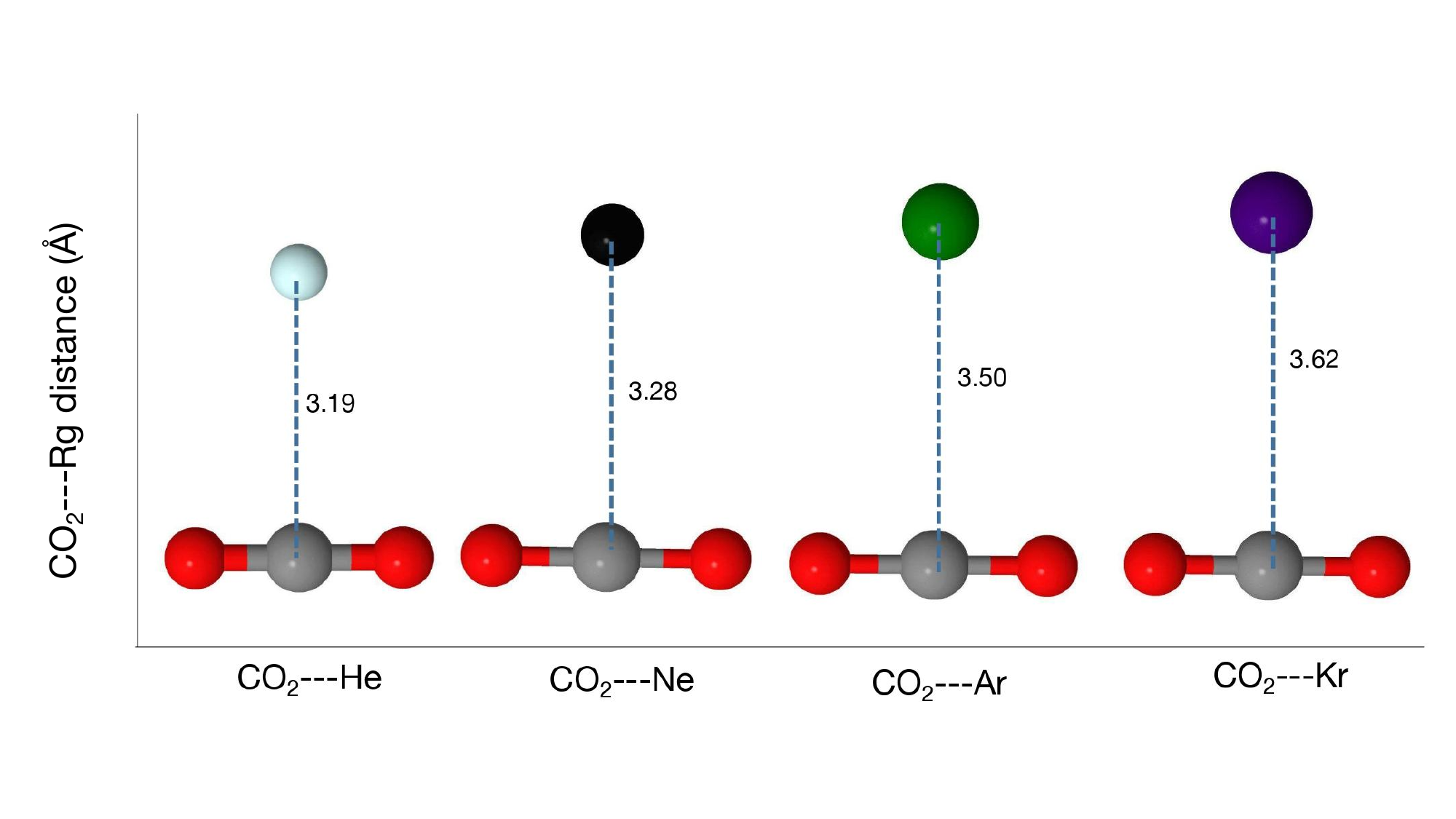}
    \caption{Distances of the Rg atoms and \ce{CO2} moiety in the \ce{CO2 \cdots Rg} complexes. The C=O distance is kept constant at 1.16 \AA.}
    \label{fig:co2_rg_geom}
\end{figure}

\FloatBarrier
\begin{table*}[ht!]
\centering
\caption{$z$-component of the response dipole moments (in D) of the \ce{CO2 \cdots Rg} complexes with percentage error (\%) relative to the response CCSD(T) values in parentheses.
FT denotes the fully converged freeze-and-thaw self-consistent embedding scheme, whereas non-FT corresponds to calculations using a static embedding potential obtained from the isolated subsystems.}
\label{tab:dipole_comparison}
\begin{tabular}{lcccc} \hline
Method / aug-cc-pVDZ & \ce{CO2\cdots He} & \ce{CO2 \cdots Ne} & \ce{CO2\cdots Ar} & \ce{CO2 \cdots Kr} \\ \hline
& & & & \\

pCCD-in-pCCD(non-FT) 
& 0.0144 (-18.18) & 0.0196 (-26.59) & 0.0720 (-2.83) & 0.0920 (-2.54) \\

fpLCCSD-in-pCCD(non-FT) 
& 0.0150 (-14.77) & 0.0228 (-14.61) & 0.0747 (0.81)  & 0.0939 (-0.53) \\

& & & & \\

pCCD-in-pCCD(FT) 
& 0.0156 (-11.36) & 0.0217 (-18.73) & 0.0772 (4.18)  & 0.0978 (3.60) \\

fpLCCSD-in-pCCD(FT) 
& 0.0160 (-9.09)  & 0.0237 (-11.24) & 0.0802 (8.23)  & 0.1001 (6.04) \\

& & & & \\

pCCD(supra.) 
& 0.0198 (12.50) & 0.0312 (16.85) & 0.0742 (0.13) & 0.0920 (-2.54) \\

fpLCCSD(supra.) 
& 0.0264 (50.00) & 0.0373 (39.70) & 0.0997 (34.55) & 0.1327 (40.57) \\

\hline 
& & & & \\

CCSD(T) 
& 0.0176 & 0.0267 & 0.0741 & 0.0944 \\ \hline
\end{tabular}
\end{table*}

\begin{table*}[ht!]
\centering
\caption{$z$-component of the expectation value (X) dipole moments (in D) of the \ce{CO2 \cdots Rg} complexes with percentage error (\%) relative to the response CCSD(T) values in parentheses.
FT denotes the fully converged freeze-and-thaw self-consistent embedding scheme, whereas non-FT corresponds to calculations using a static embedding potential obtained from the isolated subsystems.}
\label{tab:expec_dipole_comparison}
\begin{tabular}{lcccc} \hline
Method / aug-cc-pVDZ& \ce{CO2 \cdots He} & \ce{CO2 \cdots Ne} & \ce{CO2 \cdots Ar} & \ce{CO2 \cdots Kr} \\ \hline
& & & & \\
XpCCD-in-pCCD(non-FT) 
& 0.0144 (-18.18) & 0.0196 (7.49)  & 0.0720 (-2.83) & 0.0920 (-2.54) \\

XfpCCSD-in-pCCD(non-FT) 
& 0.0149 (-15.34) & 0.0220 (-17.60) & 0.0744 (0.40)  & 0.0938 (-0.64) \\

XfpLCCSD-in-pCCD(non-FT) 
& 0.0149 (-15.34) & 0.0225 (-15.73) & 0.0747 (0.81)  & 0.0940 (-0.42) \\

XCCSD-in-pCCD (non-FT) 
& 0.0151 (-14.20) & 0.0226 (-15.36) & 0.0741 (0.00)  & 0.0934 (-1.06) \\

& & & & \\

XpCCD-in-pCCD(FT) 
& 0.0156 (-11.36) & 0.0217 (-18.73) & 0.0772 (4.18)  & 0.0978 (3.60) \\

XfpCCSD-in-pCCD(FT) 
& 0.0160 (-9.09)  & 0.0234 (-12.36) & 0.0797 (7.56)  & 0.0998 (5.72) \\

XfpLCCSD-in-pCCD(FT) 
& 0.0160 (-9.09)  & 0.0236 (-11.61) & 0.0801 (8.10)  & 0.1001 (6.04) \\

XCCSD-in-pCCD (FT) 
& 0.0161 (-8.52)  & 0.0236 (-11.61) & 0.0797 (7.56)  & 0.0940 (-0.42) \\

& & & & \\

XpCCD(supra.) 
& 0.0198 (12.50)  & 0.0312 (16.85) & 0.0788 (6.34)  & 0.0920 (-2.54) \\

XfpCCSD(supra.) 
& 0.0202 (14.77)  & 0.0330 (23.60) & 0.0771 (4.05)  & 0.1108 (17.37) \\

XfpLCCSD(supra.) 
& 0.0200 (13.64)  & 0.0324 (21.35) & 0.0746 (0.67)  & 0.1094 (15.90) \\ 

XCCSD(supra.) 
& 0.0180 (2.27)   & 0.0279 (4.49)  & 0.0757 (2.16)  & 0.0966 (2.33) \\ \hline
& & & & \\
CCSD(T) 
& 0.0176 & 0.0267 & 0.0741 & 0.0944 \\ \hline
\end{tabular}
\end{table*}

The dipole moment of the van der Waals complexes of \ce{CO2 \cdots Rg} (Rg = He, Ne, Ar, and Kr) is mainly governed by electrostatic and Pauli interactions between the \ce{CO2} moiety and the inert gas atom.
The weak interactions make the dipole moments of such complexes highly sensitive to changes in electron density, providing an excellent testing ground for embedding approaches.~\cite{wesolowski1998density, wesolowski2002intermolecular, pccd-dipole-moments-jctc-2024} 
All four complexes adopt a T-shaped equilibrium  (as shown in Figure~\ref{fig:co2_rg_geom}), with the $\ce{CO2}$ molecule lying along the $  x  $-axis (carbon atom at the origin) and the rare-gas atom located along the positive $  z  $-axis. Consequently, the $  z  $-component of the dipole moment is the dominant contribution and serves as the most relevant quantity for assessing the quality of the embedding scheme.

Supramolecular dipole moments, obtained from the response density matrices of pCCD and fpLCCSD, are compared with CCSD(T) reference values in Table~\ref{tab:dipole_comparison}.
Overall, pCCD shows good agreement with the CCSD(T) benchmarks.
For the \ce{CO2 \cdots He} and \ce{CO2 \cdots Ne} complexes, pCCD slightly overestimates the dipole moments by approximately 13\%.
In the \ce{CO2 \cdots Kr} complex, the agreement improves markedly, with pCCD underestimating the dipole moment by only 2.54\%.
In contrast, fpLCCSD substantially overestimates the dipole moments across all complexes, with relative errors ranging from 35\% to 50\%.
This poor performance of fpLCCSD is not unexpected.
As shown in our previous work,~\cite{pccd-dipole-moments-jctc-2024, pccd-expectation-value-1dm-jpca-2025} the linearized correction model has difficulty accurately reproducing dipole moments for non-ionic systems, particularly small carbon-based molecules.
To further investigate this behavior, we performed the same calculations using expectation-value density matrices.
The corresponding results are summarized in Table~\ref{tab:expec_dipole_comparison}.
A comparison of the supramolecular dipole moments in Tables~\ref{tab:dipole_comparison} and~\ref{tab:expec_dipole_comparison} reveals a clear improvement in the fpLCCSD results when the expectation-value-based approach is employed (denoted as XpCCD, XCCSD, and XfpCC(S)D).
However, a detailed analysis of this lies beyond the scope of the present work, since our primary focus is the development and assessment of the pCCD-based embedding schemes.

We next examine the embedding results presented in Tables~\ref{tab:dipole_comparison} and~\ref{tab:expec_dipole_comparison}.
Given that the investigated complexes are only weakly bound, self-consistent polarization is expected to have only a minor influence on the molecular properties.
As shown in both tables, the non-FT and FT dipole moments are very similar at each level of theory (pCCD or fpLCCSD).
The FT cycles provide a slight improvement in the embedded dipole moments for both (X)pCCD and (X)fpLCC for the \ce{CO2 \cdots He} and \ce{CO2 \cdots Ne} complexes.
For the heavier \ce{CO2 \cdots Ar} and \ce{CO2 \cdots Kr} complexes, the non-FT approach yields more reliable results.
Overall, the (X)fpLCCSD-based embedding schemes produce dipole moments in closer agreement with the supramolecular CCSD(T) reference than the (X)pCCD-based schemes.
In these weakly interacting complexes, the dipole moment resides predominantly on the inert gas atom, since the interaction is too weak to significantly polarize the \ce{CO2} unit.
Consequently, the dynamic correlation correction primarily affects the dipole moment of the inert gas atom.
Here, (X)fpLCC(S)D captures electron correlation in a more balanced manner than in the supramolecular calculation, where it tends to overestimate the induced polarization (distortion) of the \ce{CO2} electron cloud.
As a result, the (X)fpLCCSD-in-pCCD embedding approaches consistently improve upon the supramolecular (X)fpLCCSD dipole moments.
Finally, among the embedding schemes considered, XfpLCCSD-in-pCCD(FT) provides the most reliable results for the \ce{CO2 \cdots He} and \ce{CO2 \cdots Ne} complexes (with dipole moment errors below 9\% and 12\%, respectively), while XfpLCCSD-in-pCCD(non-FT) yields the best agreement for the \ce{CO2 \cdots Ar} and \ce{CO2 \cdots Kr} complexes (with errors below 1\%).
It is to be noted here that for these expectation-value-based calculations, only the dipole moments were calculated using the 1-RDM obtained from cluster amplitudes. 
The pCCD electron densities, for the embedding potentials, were still obtained with the associated response $\Lambda$-equations. 
\subsection{Vertical excitation energies of the \ce{H2O \cdots NH3} complex}

\begin{figure}[h!]
    \centering
    \includegraphics[width=0.25\linewidth]{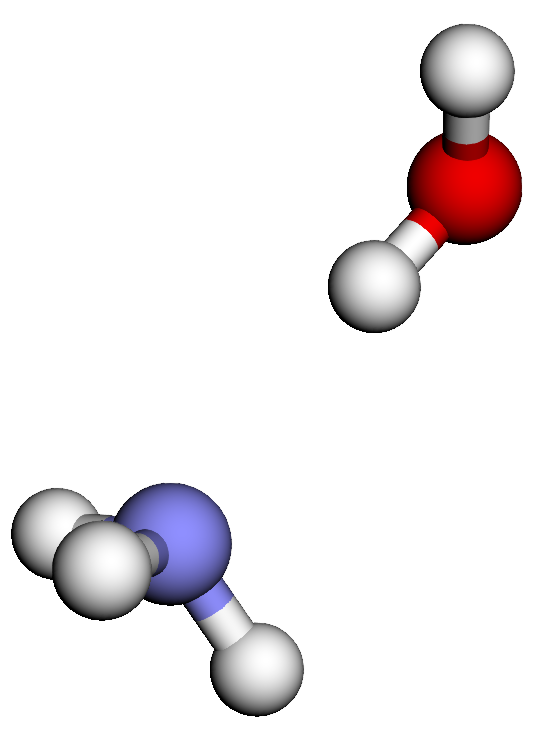}
    \caption{Schematic structure of the \ce{H2O \cdots NH3} complex.}
    \label{fig:water_ammonia}
\end{figure}

\begin{table}[!ht]
\small
\caption{Vertical excitation energies (in eV) computed with EOM-fpLCCSD/aug-cc-pVDZ for the \ce{H2O \cdots NH3} model system.
Results are shown for the isolated subsystems at the complex geometry (Iso. subsystem), the EOM-fpLCCSD-in-pCCD/aug-cc-pVDZ embedding calculation, and the supramolecular EOM-fpLCCSD/aug-cc-pVDZ reference.}
\label{tab:water-ammonia}
\centering
\begin{tabular*}{0.49\textwidth}{@{\extracolsep{\fill}}l c cc c}
\hline\hline
\shortstack{\\Source\\subsystem} & \shortstack{\\Iso.\\subsystem} & \multicolumn{2}{c}{EOM-fpLCCSD-in-pCCD} & Supramolecule \\ \hline
   &    & non-FT & FT & \\ 
\cline{3-4}
&6.43 & 6.88 &6.99  & 6.90  \\ 
~~~~~\ce{NH3}  &7.97 & 8.45 &8.58  & 8.47  \\ 
&7.97 & 8.50 &8.62  & 8.52  \\ \hline
& & & &\\
~~~~~\ce{H2O} &7.58 &7.47 & 7.44 &7.49 \\

\hline\hline
\end{tabular*}
\end{table}
Table~\ref{tab:water-ammonia} reports the three lowest-lying vertical singlet-singlet excitation energies (VEEs) originating from the \ce{NH3} moiety and one excitation from the \ce{H2O} moiety in the \ce{H2O \cdots NH3} complex (structure shown in Figure~\ref{fig:water_ammonia}).
Supramolecular EOM-fpLCCSD calculations confirm that these excitations are entirely localized on their respective parent fragments.
The remaining excitations from both fragments exhibit charge-transfer character and, therefore, cannot be described by our embedding model.
The non-FT embedding results (\ce{NH3} embedded in \ce{H2O}) are already in excellent agreement with the supramolecular reference, with deviations of only 0.02~eV.
In contrast, the FT cycles lead to a non-negligible overestimation of the VEEs for the \ce{NH3}-localized excitations (by 0.12--0.16~eV).
These numerical results suggest that the FT procedure overestimates the polarization of the \ce{NH3} ground state by the \ce{H2O} environment.
For comparison, the bare \ce{NH3} fragment at the complex geometry yields VEEs of 6.43~eV and 7.97~eV. The hydrogen bond with \ce{H2O} induces a blue shift of these transitions, which is overestimated in the FT calculations due to excessive polarization of the \ce{NH3} fragment by the environment.
These findings are consistent with previous CC-in-CC freeze-thaw (FT) calculations on the same complex.~\cite{hofener2012_cc-in-cc}

The overpolarization effect is less pronounced when \ce{H2O} is embedded in \ce{NH3}.
The first excitation of the bare \ce{H2O} fragment at the complex geometry is located at 7.58~eV.
The influence of the \ce{NH3} environment on this transition is significantly smaller than in the reverse case.
Both the non-FT and FT embedding schemes reproduce this excitation accurately, with deviations from the supramolecular reference of only 0.02~eV and 0.05~eV, respectively.

\subsection{Vertical excitation energies of microsolvated Uracil}

\begin{figure}[h!]
    \centering
    \includegraphics[width=0.65\linewidth]{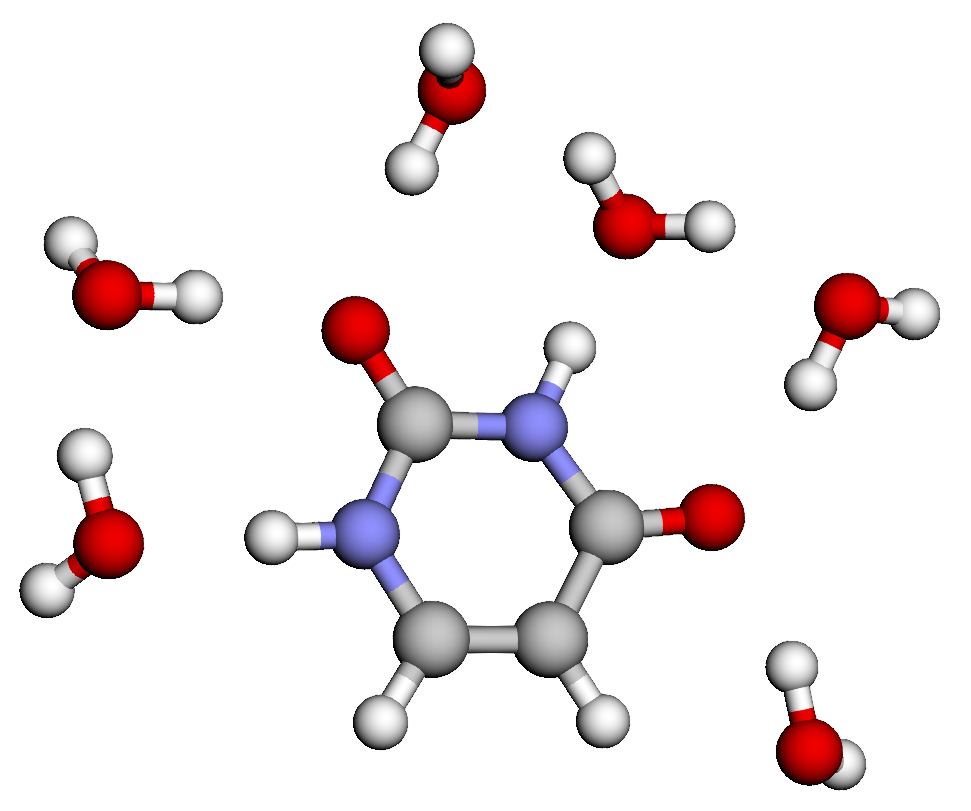}
    \caption{Schematic structure of uracil (\ce{C4H4N2O2}) surrounded by six \ce{H2O} molecules.}
    \label{fig:uracil_water}
\end{figure}

\begin{table}[!h]
\small
\caption{EOM-fpLCCSD/aug-cc-pVDZ vertical excitation energies (in eV) for the \ce{C4H4N2O2.(H2O)6} structure.
Comparison between the isolated uracil (at the complex geometry) and EOM-fpLCCSD-in-pCCD/aug-cc-pVDZ embedding results.}
\label{tab:Uracil}
\centering
\begin{tabular*}{0.49\textwidth}{@{\extracolsep{\fill}}l c c c}
\hline\hline
\shortstack{Iso.\\uracil} & \multicolumn{2}{c}{EOM-fpLCCSD-in-pCCD} & Supramolecule \\
\cline{2-3}
 & non-FT & FT & \\ 
\hline
5.62 & 5.59 & 5.75 & 5.73 \\ 
5.16 & 5.70 & 5.97 & 5.89 \\ 
\hline\hline
\end{tabular*}
\end{table}

Table~\ref{tab:Uracil} compares the vertical excitation energies (VEEs) of uracil surrounded by six water molecules (Figure~\ref{fig:uracil_water}) calculated at the supramolecular EOM-fpLCCSD level with those obtained using the embedding approach.
At the supramolecular level, the lowest-lying transition is of $ \pi \rightarrow \pi^* $ character at 5.73~eV, followed by an $n_{\rm O} \rightarrow \pi^*$ [$n_{\rm O}$ denoting the non-bonded electron pair of O] transition at 5.89~eV.
For isolated uracil (at the complex geometry), the order of these states is reversed: the $n_{\rm O} \rightarrow \pi^*$ excitation appears at 5.16~eV and the $\pi \rightarrow \pi^*$ excitation at 5.62~eV.
Consequently, the water environment induces a substantial blue shift of +0.73~eV for the $n_{\rm O} \rightarrow \pi^*$ transition and a much smaller shift of $+$0.11~eV for the $\pi \rightarrow \pi^*$ transition.
In the embedding calculations, all six water molecules are treated collectively as the environment, with uracil as the active subsystem.
Compared to the supramolecular reference, the non-FT embedding results deviate by 0.14~eV and 0.19~eV for the two transitions, respectively.
The FT cycles significantly improve the accuracy, reducing the errors to 0.02~eV and 0.08~eV.
The presence of several \ce{H2O} molecules in this case causes higher environmental polarization effect. 
This is reflected in the fact that, particularly for this system, 
the FT calculations took more (9, precisely) steps to converge, which is higher than the other weakly interacting complexes discussed above.
Hence, it is not surprising to see the improvement in the VEEs brought in by the FT cycles compared to the non-FT calculations.

\section{Conclusion}\label{sec:conclusions}

In this work, we have implemented a new WFT-in-WFT density embedding framework in the PyBEST software package.
The approach integrates the pCCD method with a simple frozen-density embedding (FDE) scheme.
Its efficiency stems from the low computational scaling of pCCD and the straightforward access to its response and expectation-value 1-RDMs, which enable inexpensive and accurate evaluation of the electron density in PyBEST.
The pCCD-based embedding scheme is particularly advantageous when DFT fails to reliably describe the ground-state density, such as in strongly correlated systems.
It accurately reproduces the dipole moments of weakly interacting \ce{CO2 \cdots Rg} complexes and the vertical excitation energies (VEEs) of microsolvated uracil. Despite the very small magnitude of the dipole moments in the \ce{CO2 \cdots Rg} systems, our embedding approach recovers them within 5--10\% of the supramolecular CCSD(T) reference, outperforming even supramolecular pCCD-based calculations.
For excited states, the method successfully describes localized excitations in the \ce{H2O \cdots NH3} complex.
The EOM-fpLCCSD-in-pCCD results yield VEEs in close agreement with supramolecular EOM-fpLCCSD calculations while correctly preserving the character and ordering of the excitations in this hydrogen-bonded system.

Overall, these results demonstrate that the proposed embedding framework can reliably treat a wide range of intermolecular interactions.
The accuracy of excited-state calculations can be further enhanced by incorporating subsystem couplings via linear-response methods.~\cite{neugebauer2007couplings, hofener2016wave, lr-pccd-jctc-2024}
The promising performance on small systems motivates extensions to more challenging cases, particularly actinide complexes embedded in large organic ligands, where pCCD has already shown reliable results for both ground and excited states.~\cite{tecmer_geminal_actinide_2015, pccd-ee-f0-actinides-pccp-2019, pccd-static-embedding}
Future developments will focus on these directions.
In addition, we plan to accelerate the embedding potential evaluation on modern GPU architectures,~\cite{pybest-gpu-jctc-2024, pybest-gpu-h100-gh200-cupy-pytorch-jctc-2026} thereby extending the applicability of the method to significantly larger systems.
\section{Acknowledgment}\label{sec:acknowledgement}

R.C. and P.T. acknowledge financial support from the OPUS research grant from the National Science Centre, Poland (Grant No. 2019/33/B/ST4/02114).
We gratefully acknowledge the Polish high-performance computing infrastructure PLGrid (HPC Centers: ACK Cyfronet AGH and WCSS) for providing computer facilities and support within the computational grant no. PLG/2025/018840.

\bibliographystyle{apsrev4-1}
\bibliography{p} 

\end{document}